\documentclass[smallextended]{svjour3}       
\smartqed  
\usepackage{graphicx}
\usepackage{amsmath,amssymb}

\usepackage{aas_macros}

\usepackage[colorlinks=true]{hyperref}

\begin{document}

\title{Observables in theories with a varying fine structure constant}


\author{A. Hees         \and
        O. Minazzoli \and
		J. Larena
}


\institute{A. Hees, J. Larena \at
              Department of Mathematics, Rhodes University \\
			  6139 Grahamstown, South Africa\\
              \email{A. Hees@ru.ac.za}           
           \and
           O. Minazzoli \at
              Centre Scientifique de Monaco \\
             UMR ARTEMIS, CNRS, University of Nice Sophia-Antipolis,\\
			Observatoire de la C\^ote d'Azur, BP4229, 06304, Nice Cedex 4, France
}

\date{Received: date / Accepted: date}

\maketitle

\begin{abstract}
We show how two seemingly different theories with a scalar multiplicative coupling to electrodynamics are actually two equivalent parametrisations of the same theory: despite some differences in the interpretation of some phenemenological aspects of the parametrisations, they lead to the same physical observables. This is illustrated on the interpretation of observations of the Cosmic Microwave Background.

\end{abstract}

\section{Introduction}
Possible variations of the fine structure constant on cosmological scales via the coupling of the electromagnetic field to Dark Energy have attracted a lot of attention  \cite{bekenstein:1982zr,sandvik:2002ly,barrow:2012wd,barrow:2013rr}. In particular, inspired by non-minimal couplings predicted by various high-energy theories (see e.g.  \cite{damour:1994fk,damour:1994uq,gasperini:2002kx,minazzoli:2014pb,peccei:1977lr,dine:1981pb,kaplan:1985nb,brax:2004fk,brax:2007xi,weltman:2008jb,ahlers:2008mq,fujii:2003fi,overduin:1997wb,minazzoli:2013fk}), many have studied such variations by introducing multiplicative couplings between the free electromagnetic Lagrangian and a scalar field (sometimes responsible for Dark Energy) \cite{damour:1994fk,damour:1994uq,gasperini:2002kx,minazzoli:2014pb,peccei:1977lr,dine:1981pb,kaplan:1985nb,calabrese:2011fr}. In this context, a series of recent papers have highlighted that such couplings may lead to non-trivial signatures in the Cosmic Microwave Background (CMB) radiation \cite{lima:1996pr,lima:2000mn,avgoustidis:2013gd,barrow:2013rr,hees:2014uq}, whereas others have claimed that such effects would not be present \cite{barrow:2014fk}. Roughly speaking, the disagreement comes from the fact that, depending on which degrees of freedom are identified with the electromagnetic field, the number of photons in the geometric optics approximation is conserved \cite{barrow:2014fk} or not \cite{lima:1996pr,lima:2000mn,avgoustidis:2013gd,barrow:2013rr,ferreira:2014al,hees:2014uq} along the propagation of the electromagnetic radiation. This paper intends to show that, in fact, at the classical level, there is no disagreement between the two points of view: if the full Lagrangian for electrodynamics is taken into account, including the part encoding the interaction with charged matter, and if this Lagrangian retains its $U(1)$ gauge invariance, then both formalisms are formally equivalent and lead to identical observable predictions. The apparent differences emerge because, depending on the choice of 4-potential, the electromagnetic field propagates differently. But this is exactly compensated by the fact that, on the other hand, this field couples to charged matter either in a non-standard way (with the presence of an explicit coupling to the scalar field) in one case, or in the standard way in the other case. Namely, in the field parametrisation leading to non-conservation of the photon number, the interaction with matter is standard, whereas, when the photon number is conserved, the presence of the coupling to the scalar field is felt locally, through the coupling to matter. Since getting any observable involves measuring the electromagnetic field via its interaction with matter, the two parametrisations lead to identical predictions.\\
In Section~II, we present the two parametrisations in a unified way and we show how to formally transform from one into the other. In Section~III, we first analyse a simple situation to explicitly show that the outcome of any measurement involving the positions and 4-velocities of charged particles is independent on the parametrisation chosen; then, we use this result to shed some light on how to reconcile the predictions of both parametrisations regarding the spectrum of the CMB. Finally, Section~IV recaps the main points of the paper.

\section{Actions, gauge invariance, field equations and geometric optic} 
In this section, we will present the action in the two parametrisations used in \cite{hees:2014uq} and \cite{barrow:2014fk}. Writing both the free part of the action and the interaction with matter, and starting with the one used in \cite{hees:2014uq}, we will derive the one used in \cite{barrow:2014fk}. We will show that the gauge transformations that leave this new action invariant ought to be modified: this modification ensures that both the new free part and the new interaction part are left invariant. We will also present the field equations and the limit of the geometric optic. The physical system considered in this study consists in an electromagnetic field described by $a^\mu$ or $A^\mu$ and a charged particle whose charge and mass are given by $q_p$ and $m_p$ respectively. We concentrate on the electrodynamics part of the action, since the discussion is not affected by the precise form of its gravitational part.

\subsection{Using the standard 4-potential}
Papers \cite{barrow:2013rr,brax:2013dq,avgoustidis:2013gd,ferreira:2014al,hees:2014uq}  preserve the standard 4-potential $a^\mu$ such that the Faraday tensor $f_{\mu\nu}$ takes its usual form:
\begin{equation}
	f_{\mu\nu}=\nabla_\mu a_\nu -\nabla_\nu a_\mu.
\end{equation}
In that case, the action is given by\footnote{Note that we use the notations of \cite{barrow:2014fk} for the coupling function but it can easily be linked to the ones used in \cite{hees:2014uq} by $e^{-2\Psi}=h(\Psi)$.}
\begin{equation}
	S  =-\frac{1}{4\mu_0 c}\int d^4x \sqrt{-g} e^{-2\Psi}  f^{\mu\nu}f_{\mu\nu} - m_p c^2 \int d\tau_p + q_p \int  a_\mu dx_p^\mu \label{eq:action1}
\end{equation}
where $c$ is the speed of light in a vacuum, $\mu_0$ is the vacuum permeability, $\tau_p$ is the particle proper time and $g$ is the determinant of the spacetime metric.

This action is invariant under standard local $U(1)$ gauge transformations:
\begin{equation}\label{eq:gauge1}
 a^\mu\rightarrow a^\mu+ \partial^\mu \chi,
\end{equation}
where $\chi$ is any function on spacetime.
It is interesting to note that the fact that the scalar field does not appear in the interaction part of the Lagrangian is required by the gauge invariance. 

Varying the action (\ref{eq:action1}) with respect to the 4-potential $a^\mu$ leads to the modified Maxwell equations
\begin{equation}\label{eq:maxwell1}
 \nabla_\mu \left(e^{-2\Psi}f^{\nu\mu}\right)=\mu_0 j_p^\nu,
\end{equation}
where $j^\nu_p$ is the 4-current defined by 
\begin{equation}\label{eq:4current}
	j^\nu_p=q_p u^\nu_p  c \gamma_p^{-1}(-g)^{-1/2} \delta^{(3)}(x^j-x^j_p(\tau_p))
\end{equation}
with $\gamma_p=dx^0/cd\tau_p$ the Lorentz factor, $u^\mu_p=dx^\mu_p/cd\tau$ the 4-velocity of the particle (such that $u^\mu u_\mu=-1$) and $x^j_p(\tau_p)$ the trajectory of the particle.

The variation of the action (\ref{eq:action1}) with respect to the position of the particle leads to the usual equations of motion
\begin{equation}\label{eq:eom1}
	m_p \frac{Du^\mu_p}{D\tau_p}=q_pf^\mu_{\ \nu}u^\nu_p \, ,
\end{equation}
where $D/D\tau$ is the covariant derivative defined (for any 4-vector $v^\alpha$) by $\frac{Dv^\alpha}{D\tau}=\frac{dv^\alpha}{d\tau}+\Gamma^\alpha_{\mu\nu}u^\mu v^\nu$ where $\Gamma^\alpha_{\mu\nu}$ are the standard Christoffel symbols.

Using the usual Lorenz gauge 
\begin{equation}
	\nabla_\mu a^\mu=0,
\end{equation}
we can write the modified Maxwell equations (\ref{eq:maxwell1}) as
\begin{equation}
	\Box a^\nu - 2 f^{\mu\nu}\partial_\mu \Psi - R^\nu_\mu a^\mu= -e^{2\Psi}\mu_0 j_p^\nu 
\end{equation}
or equivalently
\begin{equation}\label{eq:maxdev1}
	\Box a^\nu + 2 \partial^\nu a^\mu \partial_\mu \Psi - 2 \partial^\mu a^\nu \partial_\mu \Psi - R^\nu_\mu a^\mu= -e^{2\Psi}\mu_0 j_p^\nu 
\end{equation}
where $R_{\mu\nu}$ is the Ricci tensor.

\subsection{Rescaled potential}
In \cite{barrow:2014fk}, and earlier in \cite{bekenstein:1982zr}, a different 4-potential, $A_{\mu}$ was considered as the fundamental degree of freedom of the electromagnetic field:
\begin{equation}\label{eq:corr}
	A^\mu=e^{-\Psi}a^\mu,
\end{equation}
Then, one can define the Faraday tensor $F_{\mu\nu}$ as 
\begin{subequations}
\begin{eqnarray}
 F_{\mu\nu}&=&e^{-\Psi}f_{\mu\nu} \label{eq:F} \\
 &=&e^{-\Psi}\left[\partial_\mu(e^\Psi A_\nu)-\partial_\nu(e^\Psi A_\mu)\right]\label{eq:F2}\\
 &=&\partial_\mu A_\nu-\partial_\nu A_\mu +A_\nu \partial_\mu\Psi-A_\mu \partial_\nu\Psi.
\end{eqnarray}
\end{subequations}

The action (\ref{eq:action1}) with these variables reads
\begin{equation}
 S  =-\frac{1}{4\mu_0 c}\int d^4x \sqrt{-g}  F^{\mu\nu}F_{\mu\nu} - m_p c^2\int d\tau_p + q_p \int  e^\Psi A_\mu dx^\mu_p \, .\label{eq:action2}
\end{equation}

 The free part of this action is identical to the one used in \cite{barrow:2014fk}. Notice that, although the explicit coupling to the scalar field has disappeared from this free part, it has re-appeared in the interaction part.
However, the gauge transformations under which the action is invariant are now non-standard local $U(1)$ transformations that involve the scalar field:
\begin{equation}\label{eq:gauge2}
  A^\mu\rightarrow A^\mu+ e^{-\Psi}\partial^\mu\chi.
\end{equation}
This had already been noticed in \cite{bekenstein:1982zr}. It is interesting to note that each term in the action is left invariant by these gauge transformations. Indeed, the explicit coupling present in the interaction part, although it was required only in order for (\ref{eq:action2}) to be a simple re-parametrisation of (\ref{eq:action1}), also ensures that the gauge transformations leaving the free part of the action invariant keep the interaction part invariant. Retaining only the free part of action (\ref{eq:action2}) together with a standard interaction term would lead to a pathological theory that would not be gauge invariant. In this sense, the gauge invariance truly constrains the total form of the action: an action built on $A^{\mu}$ must include a non standard interaction term of the form presented here if its free part is to be related to the one in (\ref{eq:action1}) the way it is, for example, in \cite{barrow:2014fk}.   

Varying the action (\ref{eq:action2}) with respect to the 4-potential $A^\mu$ leads to the modified Maxwell equations
\begin{equation}\label{eq:maxwell2}
 \nabla_\mu \left(e^{-\Psi}F^{\nu\mu}\right)=\mu_0j_p^\nu,
\end{equation}
with $j_p^\nu$ the 4-current defined in (\ref{eq:4current}).

On the other hand, the variation of the action (\ref{eq:action2}) with respect to the position of the particle leads to the non standard equations of motion
\begin{equation}\label{eq:eom2}
	m_p \frac{Du^\mu_p}{D\tau_p}=q_p e^\Psi F^\mu_{\ \nu}u^\nu_p.
\end{equation}

The Lorenz gauge now takes the following form: 
\begin{subequations}
\begin{equation}
	\nabla_\mu (e^\Psi A^\mu)=0,
\end{equation}
or
\begin{equation}
		\nabla_\mu  A^\mu=-A^\mu \partial_\mu \Psi,
\end{equation}
and it is exactly the same gauge as the one introduced in the other parametrisation.
\end{subequations}
Using this gauge, we can write the modified Maxwell equations (\ref{eq:maxwell2}) as:
\begin{equation}
	e^{-2\Psi}\Box(e^\Psi A^\nu) -2e^{-\Psi} F^{\mu\nu} \partial_\mu \Psi - e^{-\Psi}R^\nu_\mu A^\mu=-\mu_0 j_p^\nu,
\end{equation}
or equivalently:
\begin{eqnarray}
	\Box A^\nu&+&A^\nu\Box\Psi+A^\nu \partial^\mu\Psi\partial_\mu\Psi - R^\nu_\mu A^\mu +  \label{eq:maxdev2} \\
	&&2\partial_\mu\Psi\left(\partial^\nu A^\mu+A^\mu\partial^\nu\Psi-A^\nu\partial^\mu\Psi\right)=-\mu_0 e^\Psi j_p^\nu\nonumber \; .
\end{eqnarray}

\subsection{Definition of photons}

Since the actions (\ref{eq:action1}) and (\ref{eq:action2}) are obtained by a simple change of field parametrisation, they ought to describe the same physical situation. Nevertheless, note that the inclusion of the interaction terms was crucial, since they are affected by the field redefinition. Therefore, the two sets of equations are totally equivalent. However, there are two possible different interpretations to the same set of equations --- which can naively be expected to lead to different predictions --- whether one considers $A^\nu$~\cite{bekenstein:1982zr} or $a^\nu$~\cite{sandvik:2002ly} in order to identify a photon at the classical level \footnote{Of course, for $\Psi=$constant, the two definitions describe the same field up to a change of units.}. For instance, in \cite{barrow:2014fk} the authors argue that if one considers $A^\mu$ as the ``appropriate'' fundamental electromagnetic field, then ``there will be no new observable effects on the redshift history of the cosmic microwave background radiation"; while it has previously been argued the opposite \cite{sandvik:2002ly}.

\subsubsection{Using the standard 4-potential $a^\mu$}

We can develop the Maxwell equation~(\ref{eq:maxdev1}) in a vacuum  using the geometric optics approximation. This corresponds to expanding the 4-potential $a^\mu$ as (see Eq (22.25) from \cite{misner:1973fk})
\begin{equation}
 a^\mu =\Re\left\{\left(b^\mu+\varepsilon c^\mu + \dots \right)e^{i\theta/\varepsilon} \right\}.
\end{equation}
The term proportional to $1/\varepsilon^2$ in Eq. (\ref{eq:maxdev1}) leads to the standard null geodesic equation $k_\mu k^\mu=0$ with $k_\mu = \partial_\mu\theta$ the wave vector. The next-to-leading order (proportional to $1/\varepsilon$) leads to
\begin{equation}\label{eq:b2}
 \nabla_\mu \left(b^2 k^\mu\right) =2b^2k^\mu \partial_\mu\Psi .
\end{equation}
This equation is Eq (7c) from \cite{hees:2014uq} and is also derived in Eq.~(58) from \cite{minazzoli:2014ao}. Since the number of photons is proportional to $b^2 k^0$  \cite{misner:1973fk}, in a Friedmann-Lema\^itre-Robertson-Walker (FLRW) spacetime (for illustration purposes and for the discussion of the CMB below) we have: 
\begin{equation}\label{eq:dn1}
	\dot n + 3\frac{\dot{a}}{a} n = 2 n \dot \Psi,
\end{equation}
where the dot denotes a derivative with respect to the cosmic time $t$ and $a$ is the scale factor.
\subsubsection{Using the rescaled 4-potential $A^\mu$}

Alternatively, we can develop the Maxwell equation~(\ref{eq:maxdev2}) in a vacuum using the geometric optics approximation  by expanding the 4-potential $A^\mu$ as
\begin{equation}
 A^\mu =\Re\left\{\left(B^\mu+\varepsilon C^\mu + \dots \right)e^{i\theta/\varepsilon} \right\}.
\end{equation}
The term proportional to $1/\varepsilon^2$ in Eq. (\ref{eq:maxdev2}) leads to the standard null geodesic equation $k_\mu k^\mu=0$ with $k_\mu = \partial_\mu\theta$ the wave vector. The next-to-leading order (proportional to $1/\varepsilon$) leads (after some calculations) to
\begin{equation}\label{eq:b22}
 \nabla_\mu \left(B^2 k^\mu\right) =0.
\end{equation}
We can define a new number of photons $N$, which is proportional to  $B^2 k^0$  \cite{misner:1973fk} and which is conserved. In a FLRW context, this reads:
\begin{equation}\label{eq:dn2}
	\dot N + 3\frac{\dot{a}}{a}N = 0.
\end{equation}

\subsection{Comparison between the two approaches}
It is interesting to notice that the Maxwell Equations~(\ref{eq:maxdev2}) and the equations of motion (\ref{eq:eom2}) can directly be obtained from Eqs.~(\ref{eq:maxdev1}) and (\ref{eq:eom1}) by using the transformations~(\ref{eq:corr}) and~(\ref{eq:F}).  The two formalisms seem to be completely equivalent so far since they are related by a mere field redefinition. In addition, the conservation equation (\ref{eq:dn2}) can be derived from Eq.~(\ref{eq:dn1}) by noticing that $N=e^{-2\Psi}n$. Nevertheless, here there seems to be a major difference between the two approaches:
\begin{itemize}
	\item if we use $a^\mu$: the number of photons is not conserved during the propagation of the electromagnetic signal (as seen from Eq.~(\ref{eq:dn1})), which is the result of the interaction between the scalar field and the photons. In addition, the interaction of the photons with matter is standard (as seen on Eq.~(\ref{eq:eom1})).
	\item if we use $A^\mu$: the number of photons is  conserved during the propagation of the electromagnetic signal at the geometric optic limit (as seen from Eq.~(\ref{eq:dn2})); but the interactions of the photons with matter is non standard and explicitly involves the scalar field (as seen on Eq.~(\ref{eq:eom1})).
\end{itemize}

\subsection{Fine structure constant in both representations}
In standard electromagnetism (i.e. in a theory with no scalar field), the fine structure constant is defined as
\begin{equation}
	\alpha=\frac{e^2\mu_0 c}{h},
\end{equation}
where $e$ is the electron charge and $h$ the Planck constant. It is known that this expression is modified in theories with a multiplicative coupling between the scalar field and the electromagnetic Lagrangian~\cite{bekenstein:1982zr,uzan:2011vn}.

As can be seen from the action (\ref{eq:action1}), the representation using $a^\mu$ can be interpreted as a  theory with a varying $\mu_0$ given by $\mu_0 e^{2\Psi}$. In this case, the fine structure constant is given by~\cite{bekenstein:1982zr}
\begin{equation}\label{eq:alpha}
	\alpha=e^{2\Psi}\frac{e^2\mu_0 c}{h}.
\end{equation}
On the other hand, the other representation using $A^\mu$ can be interpreted as a variation of the charge of the elementary particles. Indeed, from the action~(\ref{eq:action2}) we can interpret the representation using $A^\mu$ as a theory with varying $q_p$ given by $e^\Psi q_p$. In this case, the fine structure constant is also given by~(\ref{eq:alpha}). This shows that both representations lead to the same expression of the fine structure constant. Hence, in this type of theory, the fine structure constant can vary in space and time following the evolution of $e^{2\Psi}$ \cite{bekenstein:1982zr,sandvik:2002ly,uzan:2011vn,hees:2014uq}.

\section{Observables}
The goal of this section is to show that actual observables do not depend on the choice of the variable used to described the electromagnetic signal. In other words, we will explain why the two parametrisations of the action are completely equivalent from an observational point of view (at the classical level), which makes sense since they are simply related by a change of variables.

\subsection{A simple example}\label{sec:example}

Let us consider a situation where an electromagnetic radiation is emitted by a particle and later on interact with a detector that will for instance measure the velocity of the receiving charged particles (a current). 
This simplified situation can be modelized as follows: an electromagnetic field is emitted by a charged particle whose trajectory is given in a background space-time. This electromagnetic signal will propagate to reach another charged particle that will see its 4-velocity changed. We can neglect the backreaction of the first particle on itself and the electromagnetic signal emitted by the second particle. Moreover, we can suppose that the scalar field has a background distribution and neglect the effects of the charged particles and of the electromagnetic signal on the evolution of the scalar field (see Appendix \ref{sec:equivRepresentations} for a discussion including the whole set of equations in both parametrisations).

If we are using the formalism with $a^\mu$, the depicted situation is governed by the following system of equations
\begin{subequations}
	\begin{eqnarray}
		\Box a^\mu &+& 2 \partial^\mu a^\sigma \partial_\sigma \Psi - 2 \partial^\sigma a^\mu \partial_\sigma \Psi - R^\mu_\sigma a^\sigma= -e^{2\Psi}\mu_0 j_1^\mu \label{eq:amuevol1}\\
		m_2 \frac{Du^\mu_2}{D\tau_2}&=&q_2f^\mu_{\ \nu}u^\nu_2,	\label{eq:eom21}
	\end{eqnarray}
\end{subequations}
where the index 1 refers to the emitting particle, the index 2 refers to the receiving particle and the current $j_1^\mu$ is given by (\ref{eq:4current}).

If we are using the formalism with $A^\mu$, the same situation is governed by the following system of equations
\begin{subequations}
	\begin{eqnarray}
		\Box A^\mu&+&A^\mu\Box\Psi+A^\mu \partial^\nu\Psi\partial_\nu\Psi - R^\mu_\nu A^\nu +  \label{eq:amuevol2} \\
	 && 2\partial_\nu\Psi\left(\partial^\mu A^\nu+A^\nu\partial^\mu\Psi-A^\mu\partial^\nu\Psi\right)=-\mu_0 e^\Psi j_1^\mu\nonumber\\
	&	m_2& \frac{Du^\mu_2}{D\tau_2}=q_2 e^\Psi F^\mu_{\ \nu}u^\nu_2 \label{eq:eom22}.
	\end{eqnarray}
\end{subequations}
In these two systems of equations, $j_1^\mu(x)$ is given and the observable is given by the evolution of $u_2^\mu$. 
 The interesting point is that if $a^\mu$ is solution of the Eq.~(\ref{eq:amuevol1}), then automatically $A^\mu=e^{-\Psi}a^\mu$ is solution of the Eq.~(\ref{eq:amuevol2}). This directly implies the relation $F_{\mu\nu}=e^{-\Psi}f_{\mu\nu}$. It is then obvious that the Eqs.~(\ref{eq:eom21}) and~(\ref{eq:eom22}) are completely equivalent\footnote{See also Appendix \ref{sec:equivRepresentations}.}. As a result, the predicted evolution of the 4-velocity of the receiving particle , $u_{2}^{\mu}$, is exactly the same in both formalisms. 
This highlights that the observables predicted by the two parametrisations are exactly the same\footnote{At least as long as they depend only on the positions and 4-velocities of particles, which is a generic case.} and therefore, the two approaches are completely equivalent (at least, at the classical level). One is then free to choose the formalism used to make the calculations. Indeed, one has to keep in mind that a photon is always observed through its interaction with matter. Hence, at the classical level, it is pointless to care about which definition is supposed to describe the physical entity one calls a photon, because both definitions give rise to same observables as long as one considers the whole Lagrangian --- instead of the kinetic part only (as in \cite{barrow:2014fk}). 

\subsection{CMB temperature}
The temperature of the CMB is related to the energy density of the photons radiation \cite{durrer:2008cr}. The most suitable approach to determine the energy density of the radiation fluid is to use a microscopic approach based on the evolution of the distribution function. This approach is standard, described for example in \cite{durrer:2008cr,peter:2009kl} and is used for example in \cite{van-de-bruck:2013uq,brax:2013dq,hees:2014uq}.

If we use the parametrisation based on $a^\mu$, the non-conservation of the photon number (\ref{eq:dn1}) implies a non-conservation of the energy density of photons (see \cite{hees:2014uq} for a detailed derivation), which in a FLRW space-time takes the form:
\begin{equation}
	\dot \rho + 4 H \rho =  2 \rho \dot\Psi.
\end{equation}
This non-conservation of the energy density can directly be translated into a modification of the CMB temperature (see Eqs. (34) of \cite{hees:2014uq}). It can be physically interpreted as the fact that radiation exchanges energy with the scalar field. Moreover, the CMB spectrum does not stay Planckian in this approach and a non vanishing chemical potential is predicted (see Eq.~(35) of \cite{hees:2014uq}).\\

On the other hand, if we use the parametrisation based on $A^\mu$, the conservation of the photon number (\ref{eq:dn2}) implies a conservation of the energy density of photons, which in FLRW reads:
\begin{equation}
	\dot {\tilde \rho}+ 4H\tilde \rho = 0.
\end{equation}
The two densities defined above are related by the simple rescaling:
\begin{equation}
	\tilde\rho=e^{-2\Psi} \rho. \label{eq:rescdensities}
\end{equation}
The conservation of the energy density means that the CMB temperature follows the usual evolution (see Eq.~(20) of \cite{barrow:2013rr} or Eq.~(35) of \cite{barrow:2014fk}) and the CMB stays Planckian during the cosmological evolution. 

Therefore, there seems to be a discrepancy between the two approaches since they seem to predict two different evolutions of the CMB temperature. Nevertheless, photon numbers, CMB temperature and spectra are not observables (ie. are not measured directly). Rather, they are derived from the interactions of the CMB radiation with antennas, for example. But, we have shown in the previous section that such actual observables that involve interactions with matter are identical in both approaches. This is due to the fact that the complete sets of equations are perfectly invariant under change of parametrisation.

The apparent discrepancy comes from the fact that the transformation between the actual observable (for example a current on WMAP antenna) and the temperature of the CMB is not the same in both approaches. Indeed, Eq.~(\ref{eq:eom1}) shows that in the $a^\mu$ formalism, the interaction between photons and matter is standard. Therefore, the transformation between the observable and the temperature can be done as usual and CMB data can be used directly. On the other hand, Eq.~(\ref{eq:eom2}) shows that if we use $A^\mu$, the interaction between photons and matter is modified and explicitly involves the scalar field at the point of interaction. In this case, the transformation from the actual observable to a temperature is modified with respect to the usual treatment and it can be quite difficult to constrain the theory from this point of view. 
Despite this further complication, when analysing data in the theory considered in this paper, one has the freedom to choose between two equivalent alternatives:
\begin{itemize}
	\item To analyse the CMB temperature data in the same way as usual: the reconstruction of observables from the interaction of the CMB radiation with the measurement apparatus is standard. In this case, a deviation in the evolution of the CMB temperature is predicted by the considered model. This corresponds to using the formalism with $a^\mu$. Departure from the black-body spectrum and shifts in the temperature due to the multiplicative coupling are therefore directly constrained by the fact that the spectrum released by CMB experiments is extremely closed to Planckian (as shown in \cite{hees:2014uq}).
	\item To analyse the CMB temperature by considering non-standard interactions of radiation with matter. In this case, the theory predicts an evolution of the CMB temperature similar to the standard GR evolution. Deviations are then produced by the modification of the interaction of photons with matter. This corresponds to using the formalism with $A^\mu$. In that case, although the spectrum remains Planckian during the propagation of radiation, the multiplicative coupling is constrained by the fact that this Planckian radiation interacts non-trivially with matter at both the events of emission and measurement. 
\end{itemize}
Ultimately, as shown in the previous section, the two approaches are completely equivalent because for the same source or radiation, they will lead to identical effects on distant charges. 
In \cite{hees:2014uq}, we use CMB temperature data that has been analysed by supposing that the interaction between the photons and matter is standard. Therefore, we interpret data as a modification of the CMB temperature. Note that this approach (i.e. using $a_{\mu}$) is  simpler in practice because one does not have to model the specific way radiation interacts with matter.

\section{Conclusion}
Starting with the idea that a simple field redefinition cannot give rise to different observable predictions, we proposed that scalar multiplicative couplings to the electromagnetic Lagrangians can lead to two phenomenological predictions that are nevertheless equivalent as far as observables are concerned:
\begin{itemize}
\item On the one hand, choosing $a_{\mu}$ as the fundamental 4-potential of the electromagnetic field, the propagation of photons is affected, leading to the non-conservation of the number of photons, but the interaction of the field with matter is standard and thus, the interpretation of observations does not require a re-interpretation of actual data.
\item On the other hand, choosing $A_{\mu}$ as the fundamental 4-potential, the photon number is conserved along the propagation of the electromagnetic field, but this field interacts with matter in a non-standard way, so that, in principle, observational data should be re-analysed by taking into account the fact that the response of matter to a given electromagnetic signal is altered by the presence of the scalar field.
\end{itemize} 
This is a very similar situation to the one encountered in scalar-tensor theories with a universal conformal coupling to matter, where the coupling explicitly appears either in the gravitational part of the action (Jordan frame), or in the matter part of the action (Einstein frame): despite this apparent lack of symmetry between the two situations, observables calculated in any of the two frames are identical (see for instance \cite{flanagan:2004fk,catena:2007ij,deruelle:2011nd,chiba:2013ak,hees:2012kx}, and references therein). Also, similarly, irrespective of the level of mathematical simplicity of one parametrisation or the other, the interpretation of observations is always easier in the parametrisation for which the matter interaction remains standard, since it does not lead to any need to rederive observable quantities from first principle.\\
To conclude, representations of multiplicative couplings such as \cite{hees:2014uq} and \cite{barrow:2014fk} turn out to be formally equivalent and to lead to identical predictions, provided the interaction parts of the actions are properly taken into account. The parametrisation used in \cite{hees:2014uq} is simply more convenient because it allows a straightforward interpretation of actual data.

\begin{acknowledgements}
The authors thank J. Barrow and J. Magueijo for interesting discussions on this topic.
\end{acknowledgements}

\bibliographystyle{spphys}       
\bibliography{../../biblio_COPY} 

\appendix

\section{Equivalence of the whole set of equations}
\label{sec:equivRepresentations}
In Section~\ref{sec:example}, we show that the observables in the two representations are equivalent by considering a simplified example. In particular, we neglect the interactions of the emitting and receiving charge particles with the scalar field as well as the interactions between the EM radiation and this scalar field. This means that we treat the radiation as a test field, which is a standard hypothesis.  In this appendix, we show that this hypothesis is not required and that the two representations are formally completely equivalent. Nevertheless, in order to prove this, we need to specify the gravitational part of the action that will parametrize the evolution of the scalar field. Let us consider the following general action:
\begin{eqnarray}
S&=& \int  d^4x \sqrt{-g} \mathcal L_{\textrm{grav}}(g_{\mu\nu};\Psi)  -\int d^4 x \sqrt{-g} \frac{1}{4} e^{-2\Psi} f^2 \nonumber\\ 
&&+  \int d^4x \sqrt{-g} j^\mu a_\mu - \int d^4x \sqrt{-g} \rho \label{eq:apaction1}
\end{eqnarray}
where $g$ is the determinant of the metric tensor $g_{\mu \nu}$, $\mathcal L_{\textrm{grav}}$ is an arbitrary function of the metric tensor $g_{\mu\nu}$ (and its derivatives) and of the scalar field $\Psi$ (and its derivatives). In the case of particles, the current $j^\mu$ is given by the Eq.~(\ref{eq:4current}) and the density by 
\begin{equation}
	\rho = \sum_p m_p \gamma_p^{-1}(-g)^{-1/2}\delta^{(3)}(x^j-x^j_p(\tau_p)).
\end{equation}
Therefore, in the case of particles, the matter part of the above action is the sum of
\begin{equation}\label{eq:apactionmat1}
	S_\textrm{p}=-m_p\int d\tau_p +  q_p\int a_\mu dx^\mu_p
\end{equation}
The field equations can be derived fom (\ref{eq:apaction1}) and read as follows
\begin{subequations}\label{eq:allField1}
\begin{eqnarray}\label{field}
\frac{1}{\sqrt{-g}}\frac{\delta \left[\sqrt{-g}\mathcal L_{\textrm{grav}}\right]}{\delta g^{\mu\nu}}
     &=&  \frac{1}{2} \left( e^{-2 \Psi} T^{em} _{\mu \nu }+T^{m} _{\mu \nu }+T^{int} _{\mu \nu } \right) , \qquad\\
 \frac{\delta \mathcal L_{\textrm{grav}}}{\delta \Psi}&=& -\frac{1}{2} e^{-2 \Psi} f^2, \label{eq:Psi}
\\
\nabla_\sigma \left(e^{-2 \Psi} f^{ \mu\sigma} \right)&=& j^\mu,
\\
 \nabla_\nu  \left(e^{-2\Psi} T_{em}^{\mu\nu}\right.&+&\left. T_{m}^{\mu \nu }+T_{int}^{\mu \nu }\right)=0.
\end{eqnarray}
with
\begin{eqnarray*}
	T^{em} _{\mu \nu } &\equiv& -\frac{2}{\sqrt{-g}}\frac{\delta \left( \sqrt{-g}(-1/4~ f^2)\right)}{\delta g^{\mu \nu }}
= {f_\mu}^\sigma f_{\nu \sigma} - \frac{1}{4} g_{\mu \nu} f^2,\label{eq:defT}
\\
T^{m} _{\mu \nu } &\equiv& -\frac{2}{\sqrt{-g}}\frac{\delta \left( \sqrt{-g}(-\rho)\right)}{\delta g^{\mu \nu }}=\rho u_\mu u_\nu,\\
T^{int} _{\mu \nu } & \equiv &-\frac{2}{\sqrt{-g}}\frac{\delta \left( \sqrt{-g}j^\sigma a_\sigma\right)}{\delta g^{\mu \nu }}=-2j^\alpha a_\alpha u_\mu u_\nu.
\end{eqnarray*}
It is easy to obtain the equation of motion of point masses by varying (\ref{eq:apactionmat1})
\begin{equation}
m_p \frac{D u_p^\mu}{D\tau_p}= q_p{f^\mu}_\sigma u_p^\sigma,	
\end{equation}
\end{subequations}

In the second parametrisation the action reads
\begin{eqnarray}
S&=&\int  d^4x \sqrt{-g} \mathcal L_{\textrm{grav}}(g_{\mu\nu};\Psi) - \int d^4x\sqrt{-g}\frac{1}{4} F^2   \nonumber\\ 
&& + \int d^4x   \sqrt{-g}e^{\Psi}  j^\mu A_{\mu}  - \int d^4x \sqrt{-g} \rho . \label{eq:actiondila2} 
\end{eqnarray}
In the case of point masses, the matter part of this action becomes
\begin{equation}\label{eq:apactionmat2}
	S_p=-m_p\int d\tau_p +  q_p\int e^\Psi A_\mu dx^\mu_p
\end{equation}
The field equations therefore read as follows
\begin{subequations}\label{eq:allField2}
\begin{eqnarray}\label{field2}
\frac{1}{\sqrt{-g}}\frac{\delta \left[\sqrt{-g}\mathcal L_{\textrm{grav}}\right]}{\delta g^{\mu\nu}}
 &=&\frac{1}{2} \left( \tilde{T}^{em} _{\mu \nu }+T^{m} _{\mu \nu }+e^\Psi \tilde{T}^{int} _{\mu \nu }\right) ,  \qquad\quad  \\
 \frac{\delta \mathcal L_{\textrm{grav}}}{\delta \Psi}+\nabla_\mu \left(F^{\mu \nu}A_\nu \right) &=& -  e^\Psi  j^\mu A_\mu  , \label{eq:Psir}
\\
\nabla_\sigma \left(e^{- \Psi} F^{\mu \sigma} \right)&=& j^\mu, \label{eq:chpF}
\\
  \nabla_\nu  \left(\tilde T_{em}^{\mu\nu}\right.&+&\left. T_{m}^{\mu \nu }+ e^\Psi\tilde T_{int}^{\mu \nu }\right)=0.
\end{eqnarray}
with
\begin{eqnarray*}
\tilde{T}^{em} _{\mu \nu } &\equiv& -\frac{2}{\sqrt{-g}}\frac{\delta \left( \sqrt{-g}(-1/4~F^2)\right)}{\delta g^{\mu \nu }} = e^{-2 \Psi} T^{em} _{\mu \nu } , \quad\label{eq:deftildeT}\\
\tilde{T}^{int} _{\mu \nu } &\equiv& -\frac{2}{\sqrt{-g}}\frac{\delta \left( \sqrt{-g}j^\sigma A_\sigma\right)}{\delta g^{\mu \nu }} =e^{-\Psi}T^{int} _{\mu \nu }.
\end{eqnarray*}

The equation of motion of point masses can easily be derived by varying (\ref{eq:apactionmat2})
\begin{equation}
	m_p \frac{D u_p^\mu}{D\tau_p}= q_p e^\Psi  {F^\mu}_\nu u_p^\nu.
\end{equation}
\end{subequations}

It is straightforward to show that the set of Eqs.~(\ref{eq:allField1}) is completely equivalent to the set of Eqs.~(\ref{eq:allField2}). Indeed, all the equations from Eqs.~(\ref{eq:allField1}) can trivially be transformed into Eqs.~(\ref{eq:allField2}) by using the substitutions (\ref{eq:corr}) and (\ref{eq:F}). The only non-trivial transformation is the one related the scalar field  (\ref{eq:Psi}). However, using (\ref{eq:chpF}), one can get what follows
\begin{eqnarray*}
\nabla_\sigma \left(F^{\sigma \beta} A_\beta \right) &=& F^{\alpha \sigma} \nabla_\sigma A_\beta + A_\beta \nabla_\sigma F^{\sigma\beta}\\
&=& e^\Psi j^\mu A_\mu + F^{\alpha \sigma} \left(\nabla_\sigma A_\beta + A_\beta \nabla_\sigma \Psi\right)\\
&=&\frac{1}{2}F^2 +e^\Psi j^\mu A_\mu
\end{eqnarray*}
which, when being injected in (\ref{eq:Psir}), leads to~(\ref{eq:Psi}).
The two sets of equations are therefore perfectly equivalent. This result is obvious if one has in mind that the two sets of equations come from the same action that has only been reparametrized.

\end{document}